\documentclass[preprint,prl,aps,showpacs,footinbib,superscriptaddress,longbibliography]{revtex4-2}
\usepackage{amsmath,amssymb}
\usepackage{bm}
\usepackage{tipa}
\usepackage{upgreek}
\usepackage{comment}
\usepackage{mathrsfs}
\usepackage{graphicx}
\usepackage{braket}
\usepackage{enumitem}
\usepackage{natbib}
\usepackage{mathbbol}
\usepackage{booktabs}
\usepackage{gensymb}
\usepackage[normalem]{ulem}
\usepackage{color}
\usepackage{xcolor}
\usepackage[colorlinks,bookmarks=true,citecolor=blue,linkcolor=red,urlcolor=blue]{hyperref}
\usepackage[T1]{fontenc}
\usepackage{hyperref}

\usepackage{pifont}

\begin{document}

	\title{Probing chiral symmetry with a topological domain wall sensor}

	\author{Glenn Wagner}
	\affiliation{Department of Physics, University of Zurich, Winterthurerstrasse 190, 8057 Zurich, Switzerland}

	\author{Titus Neupert}
	\affiliation{Department of Physics, University of Zurich, Winterthurerstrasse 190, 8057 Zurich, Switzerland}
    \author{Ronny Thomale}
	\affiliation{Institut f\"{u}r Theoretische Physik und Astrophysik, 
		Julius-Maximilians-Universit\"{a}t W\"{u}rzburg, Am Hubland, 97074 W\"{u}rzburg, Germany}
   \author{Andrzej Szczerbakow} 
        \affiliation{Institute of Physics, Polish Academy of Sciences, 
        Aleja Lotnik\'ow 32/46, 02-668 Warsaw, Poland}
    \author{J\k{e}drzej Korczak}
	   \affiliation{Institute of Physics, Polish Academy of Sciences, 
        Aleja Lotnik\'ow 32/46, 02-668 Warsaw, Poland}
    \affiliation{International Research Centre MagTop, Institute of Physics, 
        Polish Academy of Sciences, Aleja Lotnikow 32/46, 02-668 Warsaw, Poland}
    \author{Tomasz Story} 
        \affiliation{Institute of Physics, Polish Academy of Sciences, 
        Aleja Lotnik\'ow 32/46, 02-668 Warsaw, Poland}
        \affiliation{International Research Centre MagTop, Institute of Physics, 
        Polish Academy of Sciences, Aleja Lotnikow 32/46, 02-668 Warsaw, Poland}
    \author{Matthias Bode} 
	\affiliation{Physikalisches Institut, Experimentelle Physik II, 
	Julius-Maximilians-Universit\"{a}t W\"{u}rzburg, Am Hubland, 97074 W\"{u}rzburg, Germany}	
    \author{Artem Odobesko}
	\altaffiliation{Corresponding author, \texttt{artem.odobesko@uni-wuerzburg.de}}
	\affiliation{Physikalisches Institut, Experimentelle Physik II, 
		Julius-Maximilians-Universit\"{a}t W\"{u}rzburg, Am Hubland, 97074 W\"{u}rzburg, Germany}
\date{\today}

\begin{abstract}
Chiral symmetry is a fundamental property with profound implications for the properties of elementary particles, that implies a spectral symmetry (i.e. $E \Rightarrow -E$ ) in their dispersion relation. In condensed matter physics, chiral symmetry is frequently associated with superconductors or materials hosting Dirac fermions such as graphene or topological insulators. There, chiral symmetry is an emergent low-energy property, accompanied by an emergent spectral symmetry. While the chiral symmetry can be broken by crystal distortion or external perturbations, the spectral symmetry frequently survives. As the presence of spectral symmetry does not necessarily imply chiral symmetry, the question arises how these two properties can be experimentally differentiated. Here, we demonstrate how a system with preserved spectral symmetry can reveal underlying broken chiral symmetry using topological defects. Our study shows that these defects induce a spectral imbalance in the Landau level spectrum, providing direct evidence of symmetry alteration at topological domain walls. Using high-resolution STM/STS we demonstrate the intricate interplay between chiral and translational symmetry which is broken at step edges in topological crystalline insulator Pb$_{1-x}$Sn$_x$Se. The chiral symmetry breaking leads to a shift in the guiding center coordinates of the Landau orbitals near the step edge, thus resulting in a distinct chiral flow of the spectral density of Landau levels. This study underscores the pivotal role of topological defects as sensitive probes for detecting hidden symmetries, offering profound insights into emergent phenomena with implications for fundamental physics.

	\end{abstract}
\maketitle

\textit{\textbf{Introduction}} --- Symmetry and observability in quantum systems embody a complicated epistemological dichotomy. As seen through Noether's theorem, symmetries may connect to conserved quantities and related operator representations, which then generically offer themselves for observability through experiment. Symmetries beyond isometries can, however, also be hidden, and rather ascribed to dynamical invariances which might not readily imply an obvious rigorous observability or determinability~\cite{RevModPhys.86.1283}. Another level of complication arises for identifying a single-particle symmetry property in a many-body phase space. Experimental signatures often assign an overall value to a many-body state, and do not allow the resolution of the individual single-particle contributions leading up to the overall many-body value. 

Such a challenging setting can appear for effective Dirac fermions in condensed matter. The electronic structure of some materials, such as graphene or topological insulators \cite{Hasan2010,Qi2011}, gives rise to well-characterized realizations of massive and massless Dirac fermions, where the fermion doubling theorem generically implies the appearance of multiple Dirac fermions~\cite{NIELSEN198120}. Leveraging the control over broken symmetries offered by condensed matter experiments, these became a testing ground for universal physical concepts related to the Dirac equation. Of the three fundamental symmetries---chiral, time-reversal, and parity---the chiral has a special status, as it implies a spectral symmetry. While the chiral symmetry relates particles and anti-particles in high-energy physics, this occurs at energies too high to be relevant for condensed matter systems. However, spectral symmetry can (re-)emerge at low energies in the condensed matter setting. While chiral symmetry implies spectral symmetry, the converse statement, i.e.\ that spectral symmetry implies chiral symmetry, is not true. 
Here, we present an example of a system in which chiral symmetry is broken in the bulk while the spectral symmetry is preserved. The question is then how one can detect the chiral symmetry breaking, i.e.\ how we can accomplish a rigorous connection between chiral symmetry and observability.

For topological systems the bulk-boundary correspondence offers a way to diagnose the state in the bulk \cite{Mong2011,Rhim2018}. The analysis can still be complicated by boundary effects at the surface. 
Therefore, we focus instead on step edges, one-dimensional defects on the two-dimensional surface of a topological insulator. 
We show that the spectrum at the step edge carries the information of the chiral symmetry, breaking in the bulk, and as such these domain walls act as sensors of the broken chiral symmetry.

\begin{figure}[t]
    \centering
    \includegraphics[width=0.55\columnwidth]{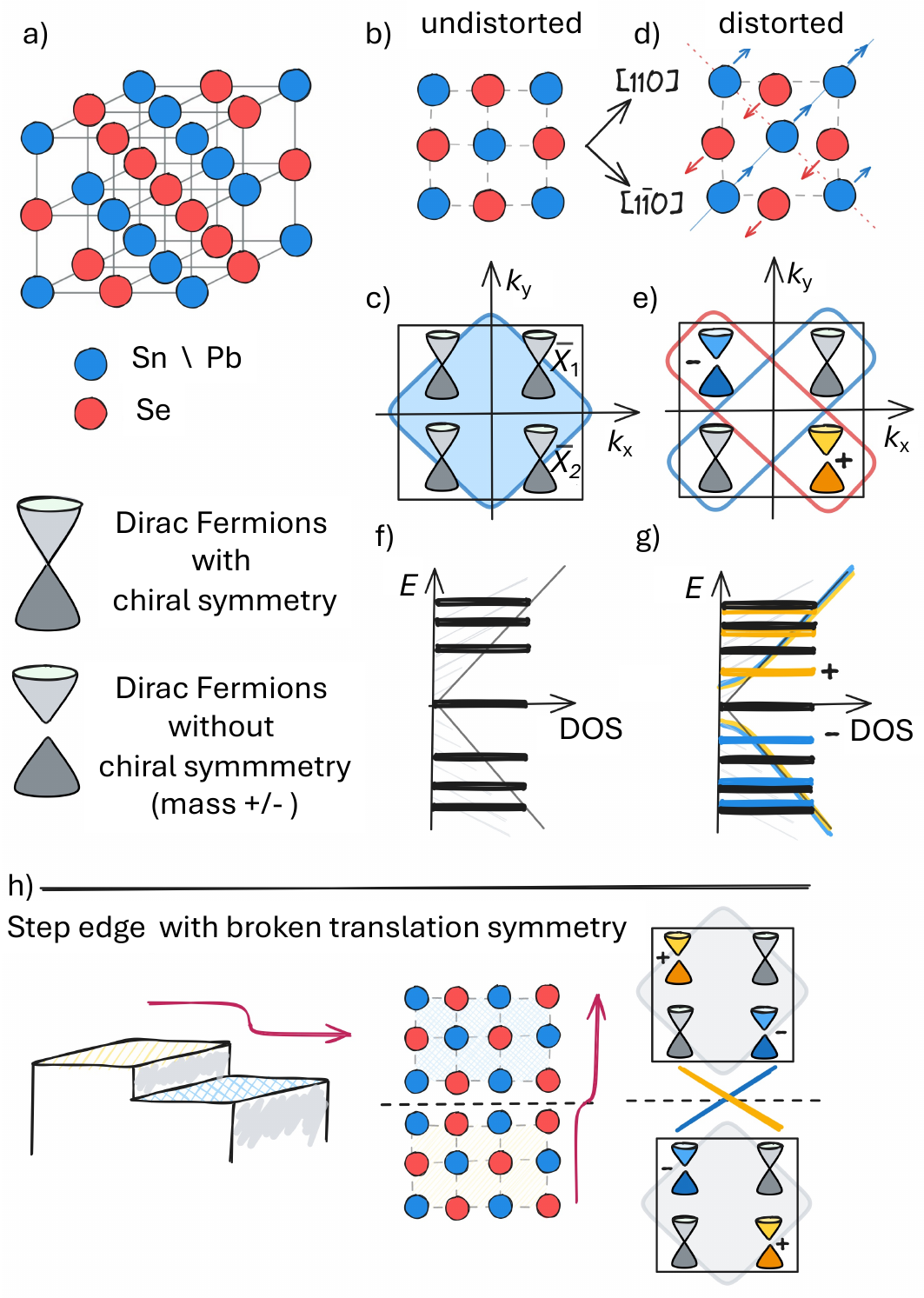}
    \caption{\textbf{Broken chiral symmetry} \textbf{a)} Crystal structure of Pb$_{1-x}$Sn$_x$Se. \textbf{b)} The top view corresponds to the undistorted cleaving [001] plane with teh schematic Brillouin zone as a blue squares and \textbf{c)} two pairs of gapless Dirac cones at $\bar{X_1}$ and $\bar{X_2}$ points.  \textbf{d)} Schematic crystal distortion of the surface with one broken mirror symmetry plane $[1\bar{1}0]$ resulting in \textbf{e)} a pair of gapped Dirac cones and a pair of gapless Dirac cones. \textbf{f)} and \textbf{g)} Corresponding DOS with overlayed LLs for a system without and with crystal distortion. \textbf{h)} Sketch of step edge with broken translation symmetry acting as a detector of the spectral flow connecting LLs of gapped Dirac cones with the same sign of the mass term}
    \label{fig_1}
\end{figure}

In our work, we illustrate this bulk-domain wall correspondence using Pb$_{1-x}$Sn$_x$Se, a topological crystalline  insulator (TCI) with rocksalt structure, see Fig.~\ref{fig_1}A. The topology of Pb$_{1-x}$Sn$_x$Se is protected by mirror symmetries \cite{Fu2011,Hsieh2012,Ando2015,Fu2011,Fang2019}. Angle-resolved photoemission spectroscopy  experiments observed a gapless surface state for Pb$_{1-x}$Sn$_x$Se \cite{Xu2012,Dziawa2012} which consists of four Dirac cones with chiral symmetry for the pristine surface, see Fig.~\ref{fig_1}B,C. However, it is known from Landau level (LL) spectroscopy that Pb$_{1-x}$Sn$_x$Se  undergoes a rhombohedral distortion which gaps out two of the four Dirac cones on the surface  (Fig.~\ref{fig_1}D,E) \cite{Okada2013}. These gapped Dirac cones appear in the spectrum as massive Landau levels, manifesting as additional peaks with the zeroth LL offset from the Dirac point and higher number LLs nearly overlapping with the LL peaks of the two remaining gapless Dirac cones, as sketched in Fig.~\ref{fig_1}F,G. 

A symmetry analysis shows that the term that opens a gap must also break the chiral symmetry of the Dirac fermions. 
 Although the chiral symmetry is broken, the spectral symmetry \( E \to -E \) remains intact in the LL spectrum. Therefore, finding a more direct signature of the chiral symmetry breaking is desirable.
Step edges in this material offer a way to resolve this question. Of particular interest are the step edges where the height changes by an odd number of half-unit cells resulting in a structural $\pi$-shift at the interface, as shown in Fig.~\ref{fig_1}H. In particular, the translational symmetry breaking from the step edge in combination with chiral symmetry breaking from the rhombohedral distortion conspire to produce a very distinct signature, i.e.\  the spectral flow density of the Landau levels close to the step edge has a ``chirality". 


In this work, we present high-resolution, low-temperature scanning tunneling spectroscopy of Landau levels on Pb$_{1-x}$Sn$_x$Se.  Our data displays the Landau levels spectral asymmetry close to the step edges offering the clearest observation of the chiral symmetry breaking in the absence of usually concomitant $E\to-E$ symmetry breaking, as probed by the step edge.

\begin{figure}[t]
    \includegraphics[width=1\columnwidth]{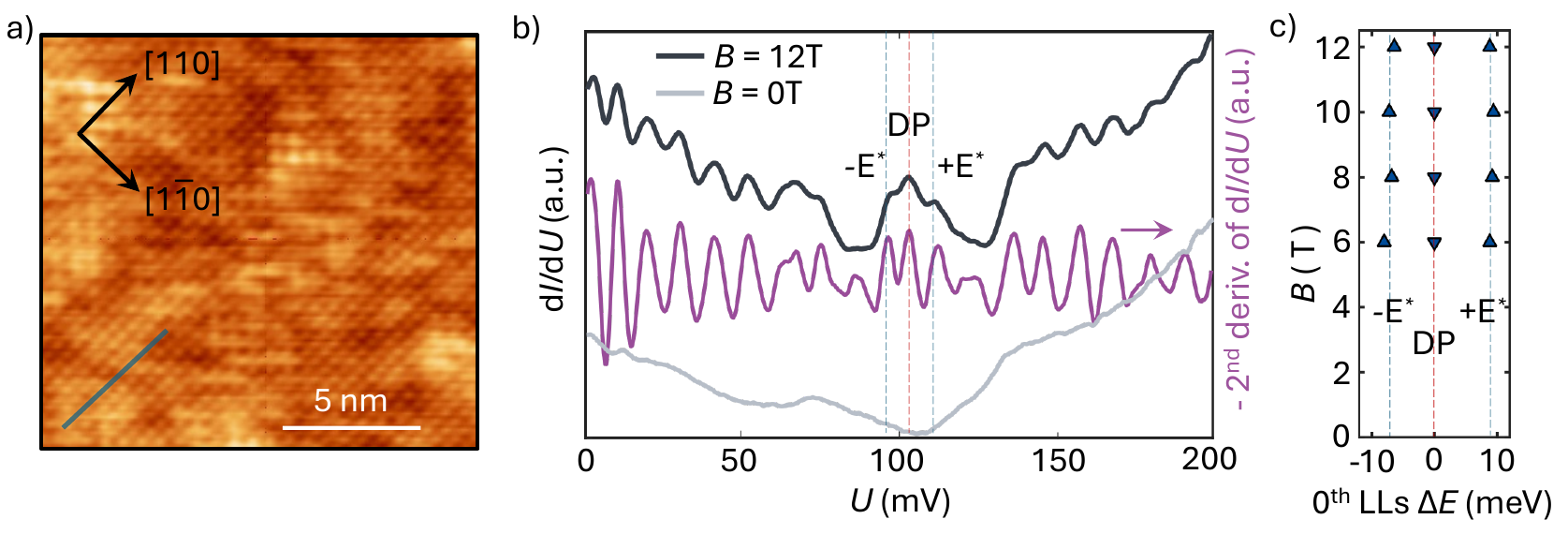}
	\caption{\textbf{Landau Levels spectrum} (\textbf{a}) Atomic resolution of a cleaved Pb$_{0.7}$Sn$_{0.3}$Se (001) surface.
		\textbf{(b)} Averaged tunneling spectra measured on a terrace along the blue line at 0T and 12T magnetic fields. In purple "--$2^{\textrm{nd}}$ derivative of $\textrm{d}I/\textrm{d}U$" signal taken at 12T.
     \textbf{(c)} Relative position of $0^{\textrm{th}}$ Landau levels  at different magnetic fields. The acquired mass  $\delta \approx 8 \pm 1$meV. 
     Setpoint parameters:  $U_{\textrm{set}} = 200\,{\rm mV}$, $I_{\textrm{set}} = 200\,{\rm pA}$, $U_{\textrm{mod}} = 1\,{\rm mV}$.}
	\label{fig_2}
\end{figure}

\textit{\textbf{Results}} ---
The Pb$_{1-x}$Sn$_{x}$Se monocrystals were grown by self-selecting vapor growth (SSVG) method \cite{Dziawa2012,SB1994}.  With $x = 0.33$ the crystals are safely in the topological regime. Samples are cleaved in ultrahigh vacuum at room-temperature. All measurements are obtained at 1.4~K.  Pb$_{1-x}$Sn$_x$Se cleaves along the [001] plane resulting in a square lattice with inter-atomic distance of $4.3$~{\AA}, indicating either the Sn/Pb or the Se sublattices, as shown in the topographic STM image in Fig.~\ref{fig_2}A. To track the density of states associated with the Dirac cone and the LLs we performed magnetic field-dependent $\text{d}I/\text{d}U$ spectroscopy.  Results for zero and 12\,T are presented in Fig.~\ref{fig_2}B as grey and black lines, respectively.  We used averaged spectra along the line shown in Fig.~\ref{fig_2}A for our analysis to minimize the contribution of inhomogeneity of the sample, caused by the random distribution of Sn/Pb atoms. A typical zero-field spectrum in this material shows a V-shaped density of states with a minimum at $\sim105$ mV corresponding to the Dirac point position of the \textit{p}-doped sample. Two shoulders are observed symmetrically at both sides of the Dirac point (at $\sim70$ mV and $\sim140$ mV) corresponding to the van Hove singularities (VHS) at the saddle point energies \cite{Okada2013}.

The averaged spectrum at the 12\,T magnetic field (Fig.~\ref{fig_2}B, black curve) shows numerous LL peaks in a wide energy range below and above the Dirac point. Comparing the zero-field spectrum with spectra measured at higher fields, an increased signal emerges precisely at the density of states minimum ($\sim 105$ meV). Obviously, it consists of three partially overlapping peaks. We present the ``--$2^{\textrm{nd}}$ derivative of $\textrm{d}I/\textrm{d}U$" signal to clarify and precisely determine the position of the $0^{\textrm{th}}$ LL peaks, see purple line in Fig.~\ref{fig_2}B.  Furthermore, two satellites labeled $\rm{E}_{–}^*$ and $\rm{E}_{+}^*$ appear on the left and right sides of the central $0^{\textrm{th}}$ LL.  As shown in Fig.~\ref{fig_2}C, the positions of these three peaks are independent of the magnetic field. We can attribute the central peak to the $0^{\textrm{th}}$ LL of massless fermions located at the Dirac point and the peaks at $\rm{E}_{-}^*$, $\rm{E}_{+}^*$ to Dirac fermions with a finite mass $m_{-/+}$ \cite{Haldane1988}. The mass acquisition arises from the rhombohedral lattice distortion that breaks the symmetry in one direction 
and opens a gap in two Dirac cones, as schematically shown in Fig.~\ref{fig_1}F,G and previously reported by Okada et al.~\cite{Okada2013}, Zeljkovic et al. \cite{Zeljkovic2015} and  Wojec et al. \cite{Wojek2015}. 

\begin{figure}[t]
    \includegraphics[width=1\columnwidth]{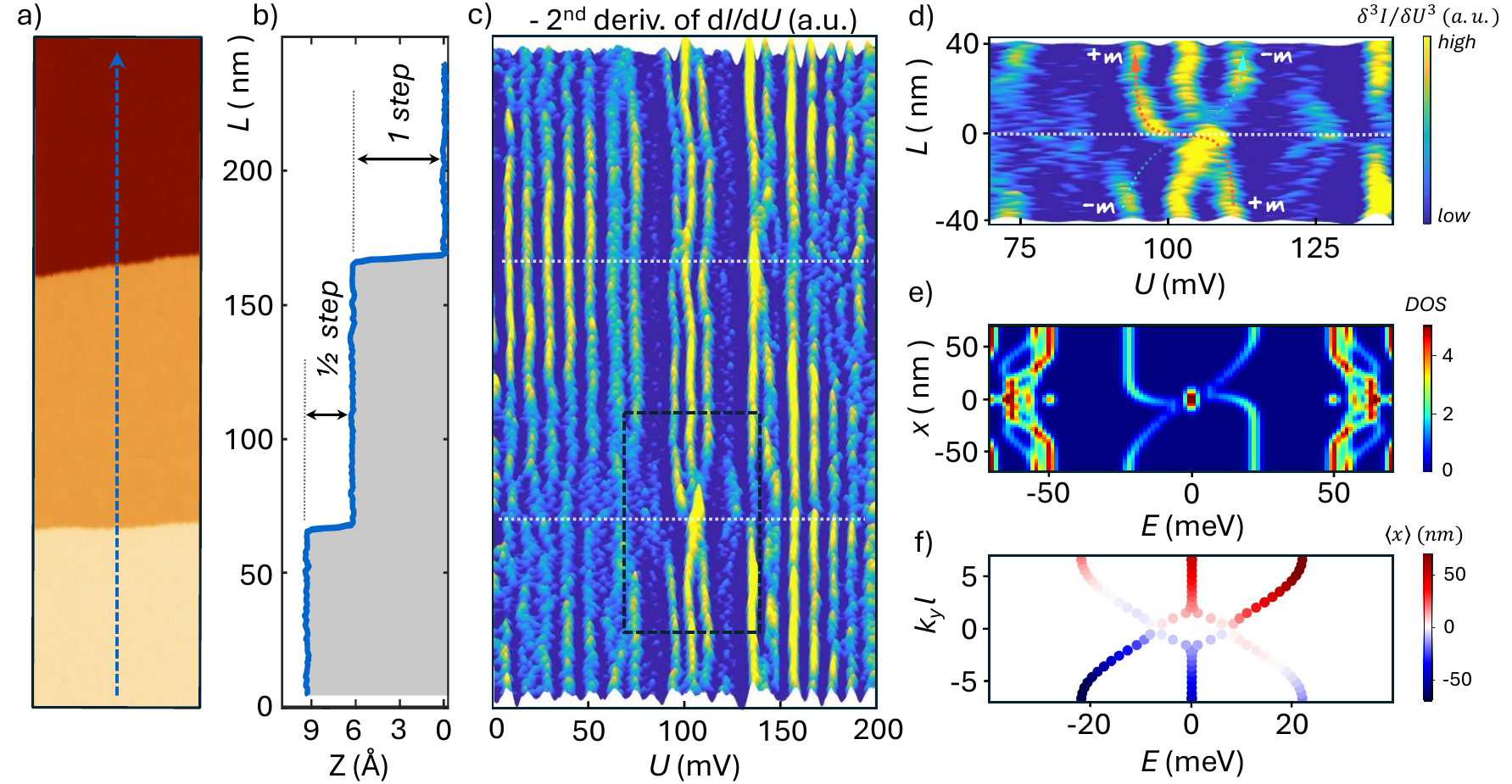}
	\caption{\textbf{LLs spectra density flow} (\textbf{a}) Topographic STM image of a Pb$_{0.7}$Sn$_{0.3}$Se surface 
		with several steps of half-- and single--unit cell height. 
		\textbf{(b)} Line profile measured along the dashed line in (a).
         \textbf{(c)} The flow of spectral density along the line is depicted by the second derivative of the \(\textrm{d}I/\textrm{d}U\) signal.
         \textbf{(d)} Zoom in the area near a half-unit cell height step showing the chiral spectral flow of $0^{\textrm{th}}$ Landau levels, attributed to symmetry breaking. \textbf{(e)} Local density of states across the step edge computed for the theoretical model in Eqns.~\eqref{eq:Ham1} and \eqref{eq:Ham2}.
		\textbf{(f)} The guiding center coordinate (given by $k_y\ell^2$ where $\ell$ is the magnetic length) is plotted against energy. The expectation value of the position $\langle x \rangle$ is encoded in the color. Away from the step edge, $\langle x \rangle$ is shifted in opposite directions at $E$ and $-E$ compared to the guiding centre coordinate. This leads to the asymmetric DOS in (e). 
     Setpoint parameters:  $U_{\textrm{set}} = 200\,{\rm mV}$, $I_{\textrm{set}} = 200\,{\rm pA}$, $U_{\textrm{mod}} = 1\,{\rm mV}$.}
	\label{fig_3}
\end{figure}

In Fig.~\ref{fig_3}, we present experimental data and a theoretical analysis revealing the spectral density flow of LLs across different step edges on Pb$_{1-x}$Sn$_x$Se. The $\textrm{d}I/\textrm{d}U$ spectra were measured along the blue hatched arrow in Fig.~\ref{fig_3}A on a Pb$_{1-x}$Sn$_x$Se surface which exhibits three atomically flat terraces separated by two step edges. The topographic line profile of Fig.~\ref{fig_3}A shows that the bottom step edge is of half-unit cell height ($3$\,{\AA}) and is therefore expected to result in a $\pi$-shift structural symmetry breaking. In contrast, the upper step edge is of one-unit cell step height ($6$\,{\AA}) preserving structural symmetry. Correlating these structural properties with the --$2^{\textrm{nd}}$ derivative of $\textrm{d}I/\textrm{d}U$ signal in Fig.~\ref{fig_3}C reveals that the structural symmetry of the step edges has a striking influence on the energetical position of the massive Dirac fermions.  

Whereas the spectral density flow of the LLs remains continuous at the step edge of a one-unit cell height, both for the higher-number LLs and the $0^{\textrm{th}}$ LL, a completely different pattern emerges at the half-unit cell step edge. A detailed analysis indicates that all higher number LLs completely vanish at the half-unit cell step edge (See Fig.~\ref{SM_fig3} in Suppl. Mat.).  The reason is the nucleation of a one-dimensional edge mode localized along the step edge \cite{Sessi2016, Jung2021, Wagner2023}, which prevents cyclotron motion of the electrons but does not hinder the existence of Landau levels with zero-point motion. Therefore, the $0^{\textrm{th}}$ LLs are conserved and reveal a complex interweaving pattern, with the position of the massive LL peaks shifting from left to right and vice versa as they cross the step. Meanwhile, the massless LL at the Dirac point retains its position. A closer look at this region, zoomed in Fig.~\ref{fig_3}D, reveals that the spectral density flow crossing the step edge is not uniform. The spectral flow intensity from the bottom to the top of the low-energy massive LL (indicated as `-' the negative mass term in Fig. 3D) gradually fades out as it approaches the step edge over a distance of approximately 20 nm, then again gradually re-emerges on the other side at the higher energy. In contrast, the intensity of the other massive LL indicated with the positive mass term '+' remains almost constant. Such a clear `chiral' spectral flow of massive LLs peaks over the step edge appeared as a common feature observed on various \textit{n}- and \textit{p}-doped Pb$_{1-x}$Sn$_x$Se samples only at the step edges.

\textit{\textbf{Model}} --- To elucidate this behaviour we provide a theoretical framework based on $k\cdot p$ approach. Two of the four inequivalent $L$ points in the three-dimensional Brillouin zone are mapped to $X_1$ under projection to the (001) surface, while the other two are mapped to $X_2$. The Dirac fermions at the $X_1$ point can hybridize, this leads to terms with $m$ and $\delta$ in the Hamiltonian \cite{Hsieh2012,Liu2013}
\begin{equation}
    H_{X_1}(\mathbf{k})=(v_1k_xs_y-v_2k_ys_x)+m\tau_x+\delta s_x\tau_y,
\end{equation}
where $s_i\ (\tau_i)$ are the Pauli matrices associated with the spin  (valley) degrees of freedom. This model has two mirror symmetries and time reversal symmetry
\begin{align}
    M_x&=-is_x,\\
    M_y&=-i\tau_xs_y,\\
    \Theta&=is_yK.
\end{align}
Furthermore, the Hamiltonian obeys a chiral symmetry $S=s_z\tau_y$ such that $S^{-1}H(\mathbf{k})S=-H(\mathbf{k})$. Taken together, the model has four Dirac cones in the two-dimensional Brillouin zone, two close to the $X_1$ point and two close to the $X_2$ point as shown in Fig.~\ref{fig_1}D. At a step edge, the surface is displaced by a single atomic layer, which is described by the translation operator $T_{t_3}=-\tau_z$  \cite{Ham}.
From Landau level spectroscopy, it is known that the material undergoes rhombohedral distortion \cite{Okada2013}. If we have a rhombohedral distortion along the [110] direction, then 
\begin{equation}
    H_{X_1}(\mathbf{k})=(v_1k_xs_y-v_2k_ys_x)+m\tau_x+\delta s_x\tau_y+\epsilon \tau_z
    \label{eq:Ham1}
\end{equation}
and the Dirac cones at $X_1$ remain gapless. Here, $\epsilon$ sets the strength of the rhombohedral distortion and is listed in Tab.~\ref{tab:parameters_Ham}. In addition, the chiral symmetry $S$ remains preserved. For the $X_2$ cones
\begin{equation}
    H_{X_2}(\mathbf{k})=(v_1k_xs_y-v_2k_ys_x)+m\tau_x+\delta s_x\tau_y+\Delta s_z\tau_y
    \label{eq:Ham2}
\end{equation}
and a gap opens up. The chiral symmetry $S$ is broken by the term proportional to  $\Delta$ (also listed in Tab.~\ref{tab:parameters_Ham}). However even in this case we retain a spectral symmetry given by $s_y\tau_z\mathcal{K}$. All together, after taking into account the rhombohebral distortion, we have two gapless and two gapped Dirac cones, as illustrated in Fig.~\ref{fig_1}E. 

We now consider this $k\cdot p$ model in an applied external magnetic field. In a magnetic field, the electron's semiclassical motion is confined to cyclotron orbits around a guiding centre and their energy levels are quantized in Landau levels. In the Landau gauge, in which $k_y$ is a good quantum number, the electron orbitals have guiding center coordinates $k_y\ell^2$, where $\ell=1/\sqrt{eB}$ is the magnetic length. In Fig.~\ref{fig_3}E we show the density of states (DOS) computed from the Landau level spectrum of the Hamiltonians \eqref{eq:Ham1} and \eqref{eq:Ham2} in the presence of the step edge. We note that the spectrum has particle-hole symmetry away from the step edge, but the symmetry is broken close to the step edge. This asymmetry is related to the positional expectation value $\langle \hat x\rangle$ at the same guiding center coordinate $k_y\ell^2$ not being the same for the states at energy $E$ and $-E$. This shift in the guiding center coordinates is opposite for the $+0^{\textrm{th}}$LL and the $-0^{\textrm{th}}$LL as seen in Fig.~\ref{fig_3}F and this is a manifestation of the broken chiral symmetry. However, in the absence of translational symmetry breaking, the spectrum remains symmetric and the chiral symmetry breaking is not visible in the DOS, as can be seen in Fig.~\ref{fig_3}E far from the edge. The step edge renders the asymmetry visible in the LDOS and hence serves as a detector of the chiral symmetry breaking.

\textit{\textbf{Conclusion}} --- We have demonstrated that chiral and translational symmetry breaking can collude to produce a spectral asymmetry in the Landau level spectrum on the surface of a topological insulator. This offers the clearest observation of the chiral symmetry breaking associated with the rhombohedral distortion of Pb$_{1-x}$Sn$_x$Se. 
The chiral symmetry breaking leads to a shift in the guiding center coordinates of the Landau orbitals. For a translationally invariant system, this shift of the guiding center coordinates does not have any observable consequences. However, once the translational symmetry is broken by the step edge, the shift of the guiding center of the $+0^{\textrm{th}}$LL in the opposite direction to the $-0^{\textrm{th}}$LL manifests as a spectral asymmetry close to the step edge. 
Our study highlights the precise detection of emergent symmetry breaking using high-resolution techniques with relativistic fermions in condensed matter. Investigating a system in vicinity of topological defects can unveil the latent symmetries, thus revealing underlying patterns otherwise obscured from the perception.

\textit{\textbf{Acknowledgements}} --- GW acknowledges funding from the University of Zurich postdoc grant FK-23-134. AO acknowledges funding supported by Deutsche Forschungsgemeinschaft (DFG, German Research Foundation) through SFB 1170 (project C02) and W{\"u}rzburg-Dresden Cluster of Excellence on Complexity and Topology in Quantum Matter -- ct.qmat. TN acknowledges support from the Swiss National Science Foundation through a Consolidator Grant (iTQC, TMCG-2-213805). RT and TN acknowledge support from the FOR 5249 (QUAST) lead by the Deutsche Forschungsgemeinschaft (DFG, German Research Foundation), in Switzerland funded by the Swiss National Science Foundation (Project 200021E-198011). The work of JK and TS  was supported by the Foundation for Polish Science project "MagTop" no. FENG.02.01-IP.05-0028/23 co-financed by the European Union from the funds of Priority 2 of the European Funds for a Smart Economy Program 2021–2027 (FENG).

\textit{\textbf{Author contributions}} --- Samples were grown by A.S., J.K., and T.S.  ~~A.O.\ and M.B.\ conceived the experiments. A.O.\ conducted the measurements and analyzed the data. The theoretical modelling was conceived by G.W., T.N., and R.T.   ~~G.W.\ performed the $k\cdot p$ calculations. All authors discussed the results. G.W.\ and A.O.\ wrote the manuscript with input from all authors.


 	\bibliographystyle{unsrtnat}
 	\bibliography{bib.bib}

\setcounter{figure}{0}
\let\oldthefigure\thefigure
\renewcommand{\thefigure}{S\oldthefigure}

\setcounter{table}{0}
\renewcommand{\thetable}{S\arabic{table}}

\newpage
\clearpage

\begin{appendix}
\onecolumngrid
	\begin{center}\textbf{\large --- Supplementary Material ---}

\end{center}

\textit{\textbf{Model}} --- Two of the four inequivalent $L$ points in the three-dimensional Brillouin zone are mapped to $X_1$ under projection to the (001) surface, while the other two are mapped to $X_2$. The Dirac fermions at the $X_1$ point can hybridize, this leads to terms with $m$ and $\delta$ in the Hamiltonian \cite{Hsieh2012,Liu2013}
\begin{equation}
    H_{X_1}(\mathbf{k})=(v_1k_xs_y-v_2k_ys_x)+m\tau_x+\delta s_x\tau_y,
\end{equation}
where $s_i\ (\tau_i)$ are the Pauli matrices associated with the spin  (valley) degrees of freedom. In Tab.~\ref{tab:parameters_Ham} we list the values of $v_1,\ v_2,\ \delta$ in the Hamiltonian. This model has two mirror symmetries and time reversal symmetry
\begin{align}
    M_x&=-is_x\\
    M_y&=-i\tau_xs_y\\
    \Theta&=is_yK.
\end{align}
Furthermore, the Hamiltonian has a chiral symmetry $S=s_z\tau_y$ such that $S^{-1}H(\mathbf{k})S=-H(\mathbf{k})$. Taken together, the model has four Dirac cones in the two-dimensional Brillouin zone, two close to the $X_1$ point and two close to the $X_2$ point. At a step edge, the surface is displaced by a single atomic layer, which is described by the translation operator \cite{Ham}
\begin{equation}
    T_{t_3}=-\tau_z.
    \label{eq:tauz}
\end{equation}
From Landau level spectroscopy, it is known that the material undergoes rhombohedral distortion \cite{Okada2013}. If we have a rhombohedral distortion along the [110] direction, then 
\begin{equation}
    H_{X_1}(\mathbf{k})=(v_1k_xs_y-v_2k_ys_x)+m\tau_x+\delta s_x\tau_y+\epsilon \tau_z
\end{equation}
and the Dirac cones at $X_1$ remain gapless. $\epsilon$ sets the strength of the rhombohedraol distortion and is listed in Tab.~\ref{tab:parameters_Ham}. In addition, the chiral symmetry $S$ remains preserved. For the $X_2$ cones
\begin{equation}
    H_{X_2}(\mathbf{k})=(v_1k_xs_y-v_2k_ys_x)+m\tau_x+\delta s_x\tau_y+\Delta s_z\tau_y
\end{equation}
and a gap opens up. The chiral symmetry $S$ is broken by the $\Delta$ term (also listed in Tab.~\ref{tab:parameters_Ham}). However even in this case we retain a spectral symmetry given by $s_y\tau_z\mathcal{K}$.
\begin{table}
    \centering
    \begin{tabular}{p{0.25\columnwidth}p{0.25\columnwidth}p{0.25\columnwidth}p{0.25\columnwidth}}
        \ \ $v_1=v_2$ & \ \ \ $m$ & \ \ \  $\delta$ & $\epsilon=\Delta$ \\  \hline\hline
        2.60 eV \AA & 0.11 eV & -0.011 eV & 0.022 eV\\
    \end{tabular}
    \caption{Parameters of the Hamiltonian used in the numerical calculations.}
    \label{tab:parameters_Ham}
\end{table}
All together, after taking into account the rhombohebral distortion, we have two gapless and two gapped Dirac cones. 

\begin{figure}
    \centering
    \includegraphics[width=\columnwidth]{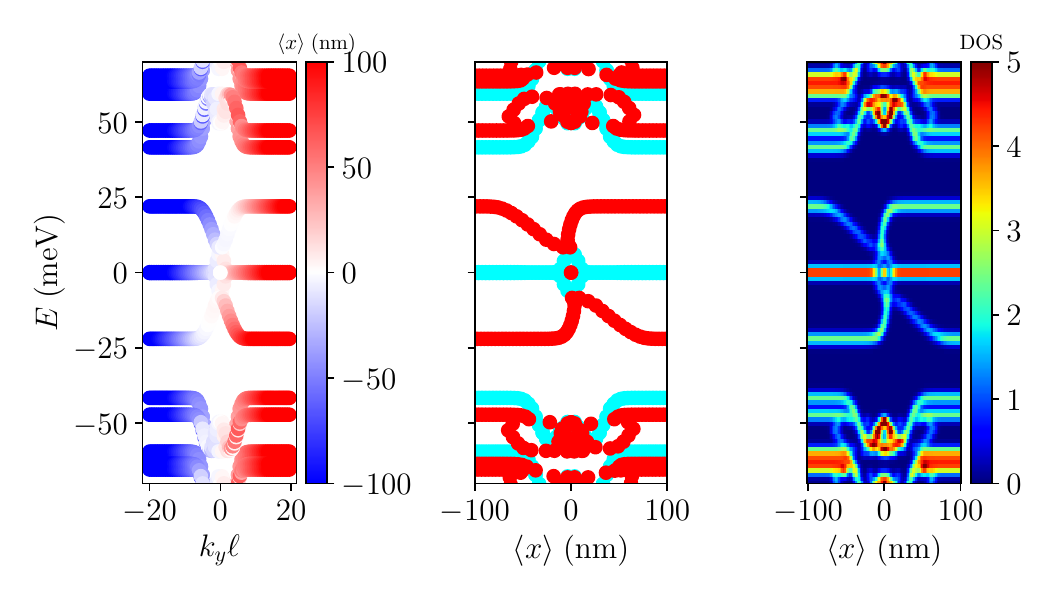}
    \caption{\textbf{Landau level spectrum.} (a) Energy spectrum as a function of the good quantum number $k_y$ with the color-coding indicating the expectation value of the position $\langle \hat x\rangle$. (b) Energy spectrum as a function of $\langle \hat x\rangle$, where the red (cyan) states are those coming from the gapped (gapless) Dirac cones. (c) Local density of states computed with smearing $\delta E=2$meV and $\delta\langle \hat x\rangle=5$nm. All subplots use a magnetic field $B=12$T and a step edge width $w=\ell_B$.}
    \label{fig:LL_numerics}
\end{figure}

We now consider this $k\cdot p$ model in an applied external magnetic field by performing a minimal coupling procedure: $k_i\to-i\hbar\partial_i-eA_i$.  In order to solve this Hamiltonian, we pick the Landau gauge $A_x=0,\ A_y=Bx$. We can introduce ladder operators such that 
\begin{align}
    \hat x&=\frac{\ell_B}{\sqrt{2}}(\hat a+\hat a^\dagger)\\
    \hat k_x&=-i\frac{1}{\sqrt{2}\ell_B}(\hat a-\hat a^\dagger)
\end{align}
where $\ell_B=\sqrt{\hbar/(eB)}$ is the magnetic length. It can be shown that 
\begin{equation}
    [\hat a^\dagger,\hat a]=1
\end{equation}
such that these ladder operators act as 
\begin{align}
    \hat a |n\rangle&=\sqrt{n}|n-1\rangle\\
    \hat a^\dagger |n\rangle&=\sqrt{(n+1)}|n+1\rangle.
\end{align}
In practice, for our numerical computations we need to truncate the occupation number basis such that we include $|n\rangle=|0\rangle,...,|N\rangle$, where we use $N=100$ to ensure convergence of the results. We can then write a finite-matrix representation of the operators $\hat a$ and $\hat a^\dagger$. We note that in the numerical calculation this truncation results in the appearance of spurious low-lying states with occupation numbers $\sim N$, which we remove from the spectrum. To model the step edge, we use the fact that the step edge corresponds to operating with Eq.~\eqref{eq:tauz} on the Hamiltonian:
\begin{equation}
    H\to \frac{1-\tanh(\hat x/w)}{2}H+\frac{1+\tanh(\hat x/w)}{2}T_{t_3}^{-1}HT_{t_3},
\end{equation}
where $w$ is the width of the step edge. In order to compute $\tanh(\hat x/w)$, we first diagonalize the operator $\hat x$, apply tanh to the eigenvalues, and then rotate back to the original basis.

The choice of gauge for this orientation of the step edge ensures that the system has translational symmetry along the $y$ direction and hence $k_y$ remains a good quantum number by which we can label the eigenstates.

In Fig.~\ref{fig:LL_numerics}(a) we show the Landau level spectrum in the presence of the step edge. We note that the spectrum has particle-hole symmetry, however the positional expectation value $\langle \hat x\rangle$ is not the same for the states at energy $E$ and $-E$. This shift in the guiding center coordinates is opposite for the +1st LL and the -1st LL and this results in the asymmetric spectrum seen in Fig.~\ref{fig:LL_numerics}(b) when the energy is plotted as a function of $\langle \hat x\rangle$. It can be seen that the gapped Dirac cones (red states) are the source of the asymmetry. This asymmetry is also visible in the density of states shown in Fig.~\ref{fig:LL_numerics}(c).

\newpage
\begin{figure}
    \centering
    \includegraphics[width=0.9\columnwidth]{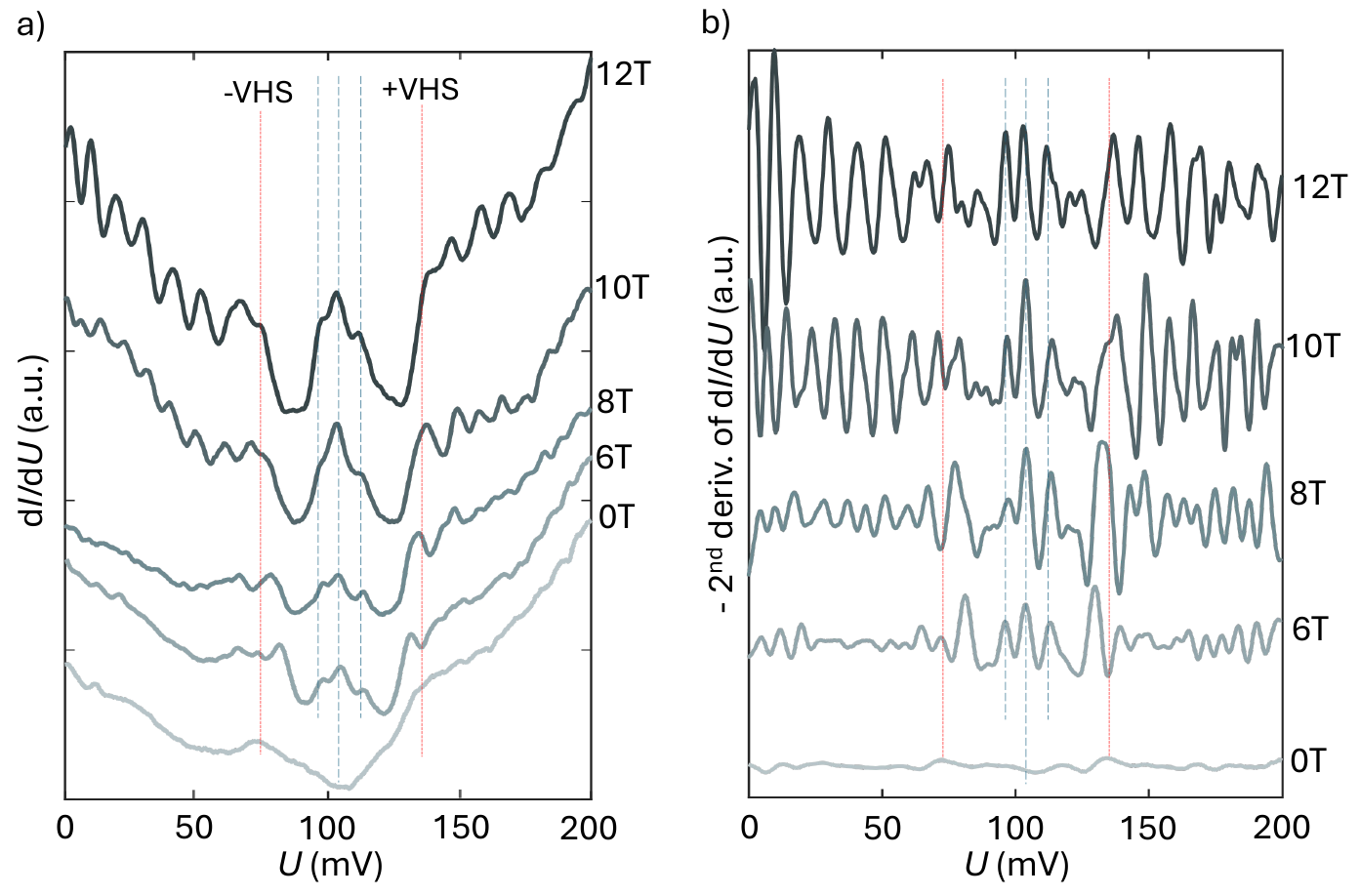}
    \caption{(\textbf{a}) Averaged spectra  along linecut presented in Fig. ~\ref{fig_2}A at various values of magnetic field strength. (\textbf{a}) Second derivative of spectra shown in (A) that were used to trace the peak positions. The estimated relative positions of  $\textrm{E}^*_{+}$ and $\textrm{E}^*_{-}$ LL peaks with respect to $0^{\textrm{th}}$ LL at the DP are used in  Fig. ~\ref{fig_2}C. The red dotted lines in (a) and (b) indicate the energies of the VHS peaks, which persist in magnetic field. Setpoint parameters:  $U_{\textrm{set}} = 200\,{\rm mV}$, $I_{\textrm{set}} = 200\,{\rm pA}$, $U_{\textrm{mod}} = 1\,{\rm mV}$.}
    \label{SM_fig2}
\end{figure}

\newpage

\begin{figure}
    \centering
    \includegraphics[width=0.9\columnwidth]{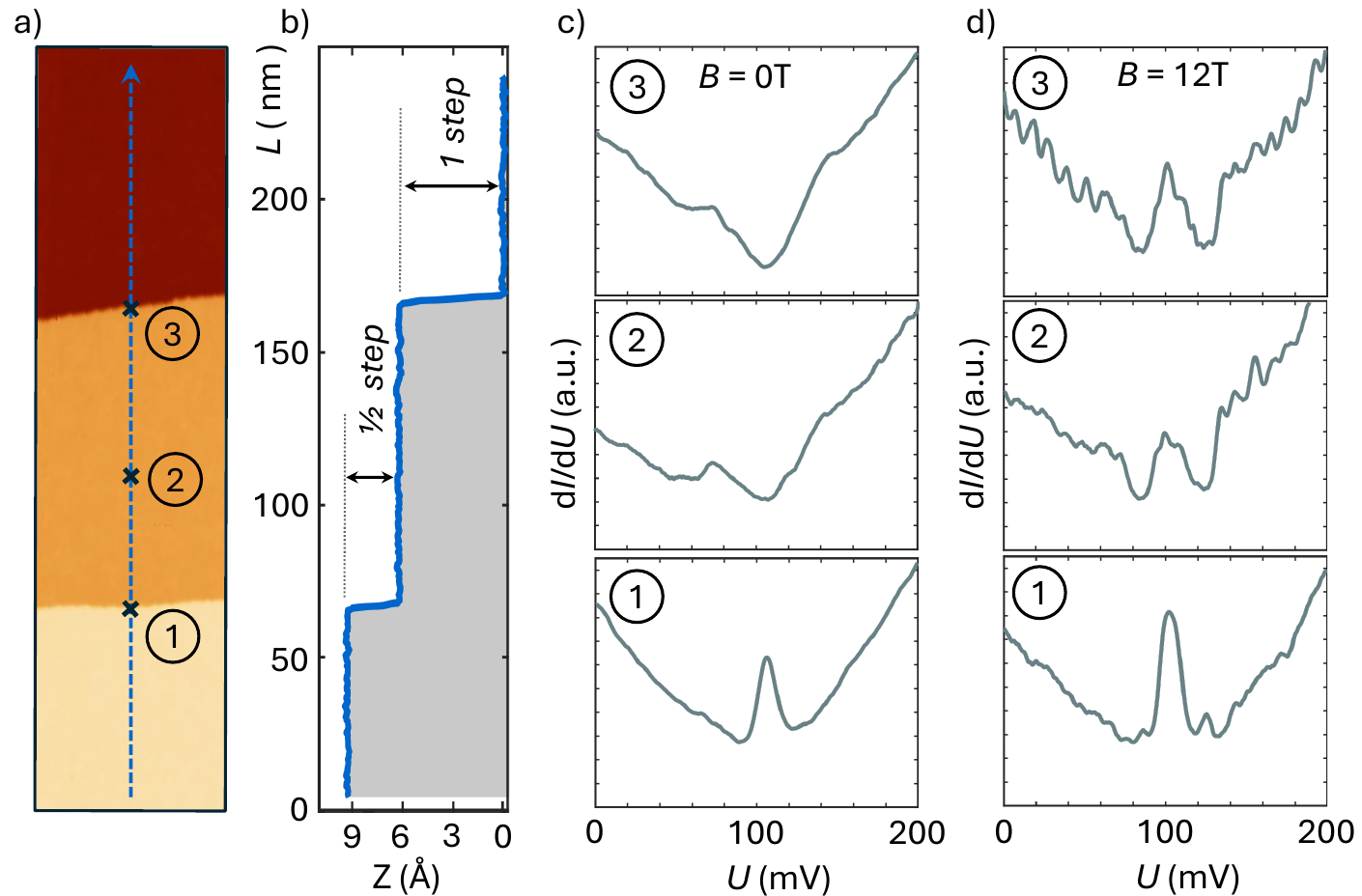}
    \caption{\textbf{a}) Topographic STM image of a Pb$_{0.7}$Sn$_{0.3}$Se surface with flat terraces separated by several step edges of half-- and single--unit cell height. Marks indicate the positions of a single point spectra measured at zero and 12T fields presented in (c) and (d) respectively. \textbf{(b)} Line profile measured along the dashed line in (a).}
    \label{SM_fig3}
\end{figure}

The zero-field point spectrum (1) in Fig.~\ref{SM_fig3}C, measured at the step with a half-unit cell height, shows a pronounced peak at the Dirac point energy ($\approx 105$~meV). This peak corresponds to a one-dimensional (1D) topological edge state with flat band dispersion which emerges at step edges with a translational symmetry $\pi$-shift \cite{Sessi2016}. Point spectrum (2), measured at the terrace far from the step edges, exhibits a dip at the Dirac point and two shoulders corresponding to van Hove singularities. Spectrum (3), taken at the step edge with a single-unit cell height that preserves translational symmetry, is similar to spectrum (2).

At higher magnetic fields (Fig.~\ref{SM_fig3}D), Landau levels appear in spectra (2) and (3) due to the quantization of cyclotron orbits. However, in spectrum (1), LLs are absent except for a peak at the Dirac point, indicating a topological 1D edge state that is robust to the magnetic field. Higher LLs require cyclotron motion and thus cannot form in a 1D system. The $0^{\textrm{th}}$ LL of Dirac fermions, however, is linked to their intrinsic properties and overlaps with the signal from the 1D edge state.

\newpage
\begin{figure}
    \centering
    \includegraphics[width=0.7\columnwidth]{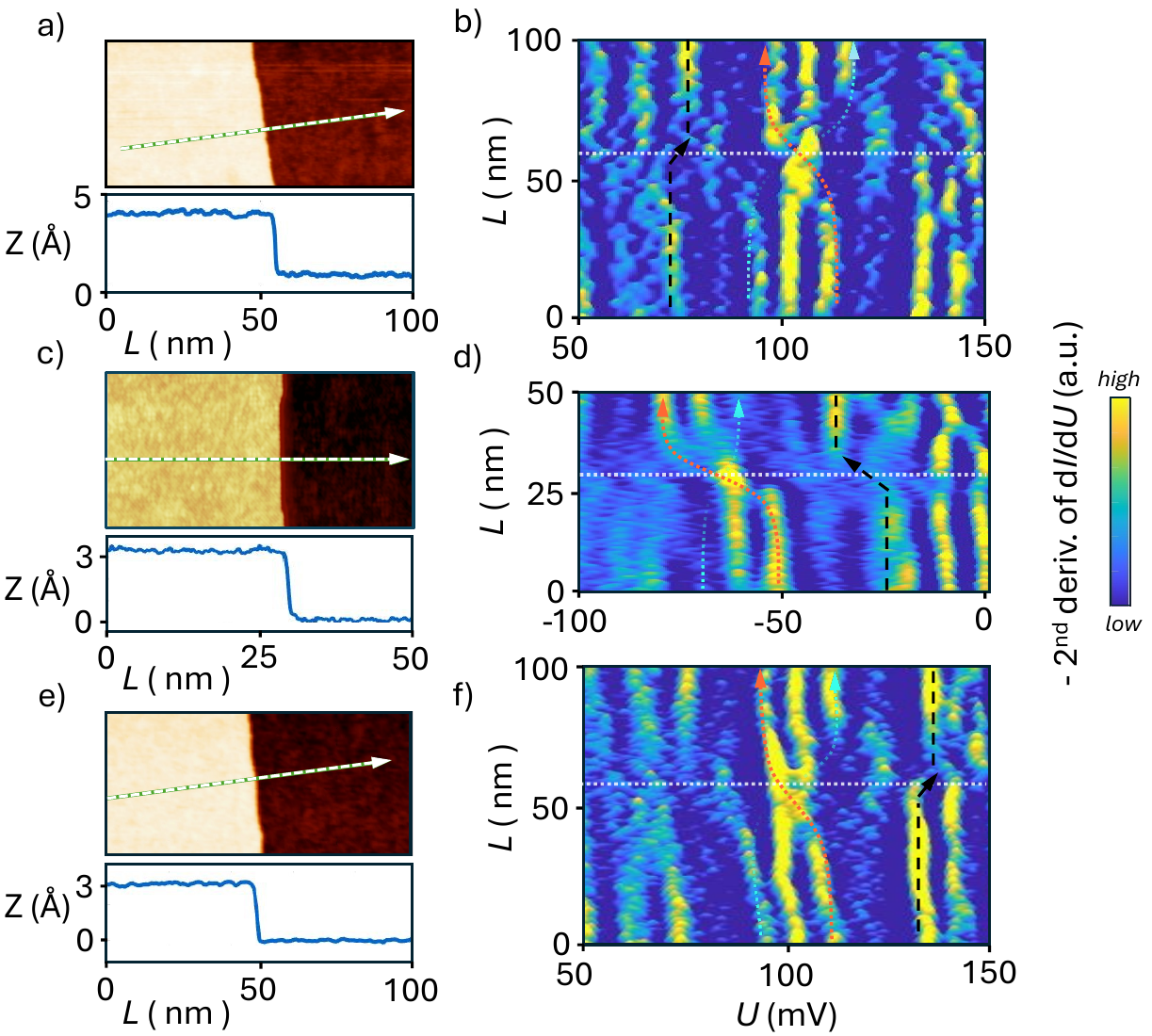}
    \caption{Spectral density flow of Landau levels across the step edge with half-unit cell height measured on the samples with different doping. (\textbf{a}) STM topography image with z-profile of \textit{p}-doped sample with Dirac point (DP) at $\approx 105$~meV and (\textbf{b}) spatial variation of --$2^{\textrm{nd}}$ derivative of $\textrm{d}I/\textrm{d}U$ signal measured along white dashed arrow in (a). The dotted line in (b) indicates the position of the edge step. (\textbf{c}) and (\textbf{d}) The same on the \textit{n}-doped sample with DP at $\approx -70$~meV. (\textbf{e}) and (\textbf{f}) The same on \textit{p}-doped sample with DP at $\approx 105$~meV}
    \label{SM_fig4}
\end{figure}

Uneven defect distribution along the Pb$_{1-x}$Sn$_{x}$Se sample causes variations in the Dirac point position relative to the Fermi level. This effect is often observed along steps, which can be tracked by energy position shifts of Landau levels. For example, in Fig.~\ref{SM_fig4}B, the LL with $n = -1$, marked by the black dashed line, slightly shifts by a few mV towards positive energy while crossing the step edge. Thus, the Dirac point position at the bottom of Fig.~\ref{SM_fig4}B is $\approx 102$~mV, while at the top it is $\approx 104$~mV. Nevertheless, the chiral character of the spectral density flow of $0^{\textrm{th}}$ LLs is preserved but experiences a slight positive energy tilt. In Fig.~\ref{SM_fig4}D, this effect is more pronounced, where the LL with $n = +1$, marked by the black dashed line, shifts significantly towards negative energy. Consequently, the Dirac point at the bottom is $\approx -60$~mV and at the top is $\approx -70$~mV. This complicates the interpretation of $0^{\textrm{th}}$ LLs crossings, but the nonuniform spectral density flow still remains evident. We observe, that shifts in LLs positions at step boundaries are very common in this material and can be positive or negative, e.g. as seen in (e,f), which is the same \textit{p}-doped sample as in (a,b), but with the Dirac point shifting towards positive energy.

\end{appendix}
\end{document}